\documentclass[conference,10pt]{IEEEtran}
\IEEEoverridecommandlockouts
\usepackage{multirow}
\usepackage{amsmath,amssymb,amsfonts,amsthm}
\usepackage{algpseudocode}
\usepackage{algorithmicx}
\usepackage{algorithm}
\usepackage{textcomp}
\usepackage{xcolor}
\usepackage{tikz}
\usepackage{mathtools}
\usetikzlibrary{calc}
\usetikzlibrary{positioning}
\usepackage{qcircuit}
\usepackage{pgfplots}
\usepackage{graphics}
\usepackage{caption}
\usepackage{subcaption}
\usepackage{smartdiagram}
\usesmartdiagramlibrary{additions}
\pgfplotsset{compat=1.16}
\usepackage[backend = biber, style=ieee, maxnames = 6, minnames = 1,url=false, eprint=false, doi=false]{biblatex}
\usepackage{stfloats}
\usepackage{hyperref}

\theoremstyle{plain}

\newtheorem{thm}{Theorem}
\newtheorem{lem}{Lemma}

\newtheorem{prob}{Problem}

\renewcommand{\epsilon}{\varepsilon}

\newcommand{\e}{\ensuremath{\mathcal{E}}}

\newcommand{\h}{\ensuremath{\mathcal{H}}}

\renewcommand{\o}{\ensuremath{\mathcal{O}}}

\newcommand{\N}{\ensuremath{\mathcal{N}}}

\renewcommand{\k}{\ensuremath{\mathcal{K}}}

\renewcommand{\u}{\ensuremath{\mathcal{U}}}

\newcommand{\tr}{{\rm tr}} 

\newcommand{\bra}[1]{\langle #1 |}
\newcommand{\ket}[1]{| #1 \rangle}

\newcommand{\ketbra}[2]{| #1 \rangle \langle #2 |}

\tikzset{every picture/.append style={remember picture}}
\tikzset{tensor/.style={rectangle, inner sep=2pt,draw,minimum size=0.5cm}}
\newcommand{\tikzdiagram}[2][]{
	\begin{array}{cc}
		\begin{tikzpicture}[baseline,#1]
			#2
		\end{tikzpicture}
	\end{array}
}

\makeatletter
\newcommand{\linebreakand}{%
  \end{@IEEEauthorhalign}
  \hfill\mbox{}\par
  \mbox{}\hfill\hspace{7pt}\begin{@IEEEauthorhalign}
}
\makeatother

\def\BibTeX{{\rm B\kern-.05em{\sc i\kern-.025em b}\kern-.08em
		T\kern-.1667em\lower.7ex\hbox{E}\kern-.125emX}}
\begin{document}
	
	\title{Approximation Algorithm for Noisy Quantum Circuit Simulation
	}
	
    \author{\IEEEauthorblockN{1\textsuperscript{st} Mingyu Huang}
    \IEEEauthorblockA{\textit{Institute of Software, Chinese Academy of Sciences} \\
    \textit{University of Chinese Academy of Sciences} \\
    Beijing, China \\
    huangmy@ios.ac.cn}
    \and
    \IEEEauthorblockN{2\textsuperscript{nd} Ji Guan}
    \IEEEauthorblockA{\textit{Institute of Software, Chinese Academy of Sciences} \\
    Beijing, China \\
    guanj@ios.ac.cn}
    \linebreakand
    \IEEEauthorblockN{3\textsuperscript{rd} Wang Fang}
    \IEEEauthorblockA{\textit{Institute of Software, Chinese Academy of Sciences} \\
    \textit{University of Chinese Academy of Sciences} \\
    Beijing, China \\
    fangw@ios.ac.cn}
    \and
    \IEEEauthorblockN{4\textsuperscript{th} Mingsheng Ying}
    \IEEEauthorblockA{\textit{Institute of Software, Chinese Academy of Sciences}\\
    \textit{Tsinghua University} \\
    Beijing, China \\
    yingms@ios.ac.cn}
    \and
    }
	
	\maketitle
	
	\begin{abstract}
		
        Simulating noisy quantum circuits is vital in designing and verifying quantum algorithms in the current NISQ (Noisy Intermediate-Scale Quantum) era, where quantum noise is unavoidable. However, it is much more inefficient than the classical counterpart because of the quantum state explosion problem (the dimension of state space is exponential in the number of qubits) and the complex (non-unitary) representation of noises. Consequently, only noisy circuits with up to about 50 qubits can be simulated approximately well. This paper introduces a novel approximation algorithm for simulating noisy quantum circuits when the noisy effectiveness is insignificant to improve the scalability of the circuits that can be simulated. The algorithm is based on a new tensor network diagram for the noisy simulation and uses the singular value decomposition to approximate the tensors of quantum noises in the diagram. The contraction of the tensor network diagram is implemented on Google's TensorNetwork. The effectiveness and utility of the algorithm are demonstrated by experimenting on a series of practical quantum circuits with realistic superconducting noise models. As a result, our algorithm can approximately simulate quantum circuits with up to 225 qubits and 20 noises (within about 1.8 hours). In particular, our method offers a  speedup over the commonly-used approximation (sampling) algorithm --- quantum trajectories method~\cite{isakov2021simulations}. Furthermore, our approach can significantly reduce the number of samples in the quantum trajectories method when the noise rate is small enough.
	\end{abstract}
	

\section{Introduction}\label{sec:intro}

Since the achievement of quantum supremacy over classical computing \cite{arute2019quantum}, quantum processors with an increasing number of quantum bits (qubits) \cite{gong2021quantum, ibm2022quantum} have been manufactured. This progress has marked the transition to the NISQ (Noisy Intermediate-Scale Quantum) era \cite{preskill2018quantum}. Even though current quantum processors have a limited number of qubits and are susceptible to quantum noise, they've been used in numerous applications, highlighting the potential benefits of NISQ devices \cite{49058, doi:10.1126/science.abb9811, 46227}. The circuits that carry out computational tasks are at the heart of these processors.

Building and testing quantum circuits in real-world environments is crucial in quantum computing. These environments often introduce \emph{noises}, a common challenge in the NISQ era. Simulating these circuits on classical computers before building them is beneficial—it saves costs and helps avoid potential issues when implementing quantum circuits and reading outputs from real devices. Consequently, it is urgently necessary to develop efficient simulation algorithms for noisy quantum circuits.

\textbf{Noiseless Simulation:} The widely used method of noiseless simulation is to calculate the transitions of the state vector through a sequence of quantum gates modeled by unitary matrices. This method is straightforward and integrated into most popular software development kits for quantum computing (e.g., Qiskit, Cirq, etc.). However, the exponential number of terms in these representations restricts the number of qubits that can be simulated. To overcome this issue, one popular method employs a data structure, the \emph{tensor network}, to capture the locality (a quantum gate is only applied on 1 or 2 qubits) and regularity (the pattern of quantum gates in practical circuits is regular) of quantum circuits. This approach has been successfully utilized in simulating large noiseless quantum circuits~\cite{haner20175,huang2020classical,li2019quantum,pednault2017breaking,villalonga2019flexible, PhysRevLett.128.030501}. An alternative approach is the Decision Diagram-based (DD-based) method~\cite{10.1109/DAC18074.2021.9586191, zulehner2018advanced}, akin to the Binary Decision Diagrams (BDD) in classical computing. This technique optimizes memory usage and performance by compactly storing all state amplitudes in a specialized data structure. While effective for certain types of circuits, particularly those that maintain manageable data structure sizes throughout the simulation, its performance diminishes with circuits that have gates with arbitrary parameters, such as those used in quantum supremacy experiments~\cite{10.1109/DAC18074.2021.9586191, hillmich2021accurate}.  

\textbf{Noisy Simulation:} For simulating noisy quantum circuits, most software packages~\cite{Qiskit, cirq_developers_2022_7465577} employ the density matrix representation. While this approach is standard, it struggles with scalability when applied to circuits with a high qubit count. To enhance the scalability of noisy circuit simulations, several approximation algorithms have been introduced. These include the quantum trajectories method, which offers a probabilistic approach to simulate noise effects~\cite{isakov2021simulations}. Algorithms such as MPS (Matrix Product State) \cite{zhou2020limits}, MPO (Matrix Product Operators) \cite{woolfe2015matrix, noh2020efficient}, and MPDO (Matrix Product Density Operators) \cite{cheng2021simulating} use tensor network representations along with SVD (singular value decomposition) for approximation. The DD-based simulation can also be extended to noisy simulation for some specific circuits and noise types~\cite{ddsim_noise}. However, challenges like the quantum state explosion problem restrict current simulations to about 50 qubits. The simulation demand of the NISQ circuits with up to hundreds of qubits cannot be satisfied.

\textbf{Contributions of This Paper:} To address this gap, we developed a new approximation algorithm for simulating noisy quantum circuits. We introduce a novel tensor network diagram for this simulation. In our diagram, noisy quantum circuits are represented by double-size tensors and can be well approximated by performing SVD on their tensor representation. Based on these, an approximation algorithm is developed and implemented with Google TensorNetwork~\cite{roberts2019tensornetwork}. The effectiveness and utility of our algorithm are confirmed by experimenting with three types of practical quantum circuits (algorithms). The experimental results show that our algorithm can approximately simulate quantum circuits with up to 225 qubits and 20 noises (within about 1.8 hours). In particular, our method offers a speedup over the commonly-used approximation (sampling) algorithm — quantum trajectories method~\cite{isakov2021simulations}.

	\section{Preliminary}\label{sec:pre}
	%
    {\bf Basics of Quantum Computation:}
	We start by recalling some basic concepts of quantum circuits. Ideally, a quantum computer without noise is a closed system. In this case, quantum data are mathematically modeled as complex unit vectors in a $2^n$-dimensional Hilbert (linear) space $\h$. Such a quantum datum is usually called a \emph{pure state} and written as $\ket{\psi}$ in the Dirac notation, and $n$ represents the number of involved  \emph{quantum bits (qubits)}. Specifically, a qubit is a quantum datum in a $2$-dimensional Hilbert space, denoted by
	$\ket{q}=\begin{pmatrix}a\\ b  \end{pmatrix}=a\ket{0}+b\ket{1}$ with $\ket{0}=\begin{pmatrix}1\\ 0\end{pmatrix}$ and $\ket{1}=\begin{pmatrix}0\\ 1\end{pmatrix}$,
	where complex numbers $a$ and $b$ satisfy the normalization condition $|a|^2+|b|^2=1$. Here, the orthonormal basis $\{\ket{0}$, $\ket{1}\}$ of the Hilbert space corresponds to the  values  $\{0,1\}$ of a bit in classical computers. A quantum computing task is implemented by a \emph{quantum circuit}, which is mathematically represented by a $2^n\times 2^n$ unitary matrix $U$, i.e., $U^\dagger U=UU^\dagger=I_n$, where $U^\dagger$ is the (entry-wise) conjugate transpose of $U$ and  $I_n$ is the identity matrix on $\h$. For an input $n$-qubit datum $\ket{\psi}$, the output of the circuit is a datum of the same size: $\ket{\psi'} = U\ket{\psi}.$
	
\begin{figure}
        \begin{align*}
        \begin{array}{c}
            \Qcircuit @C=0.5em @R=0.5em @!R {
            & \gate{\mathrm{R_Y}\,(\mathrm{\frac{-\pi}{2}})} & \gate{\mathrm{R_Z}\,(\mathrm{\frac{\pi}{2}})} & \ctrl{1} & \ctrl{1} & \gate{\mathrm{R_X}\,(\mathrm{\pi})} & \qw \\
            & \gate{\mathrm{R_Y}\,(\mathrm{\frac{-\pi}{2}})} & \gate{\mathrm{R_Z}\,(\mathrm{\frac{\pi}{2}})} & \ctrl{-1} & \gate{\mathrm{R_Z}\,(\theta)} & \gate{\mathrm{R_X}\,(\mathrm{\pi})} & \qw
            }
        \end{array}
        \end{align*}
        \caption{A 2-qubit QAOA circuit}
        \label{fig:circuit_example}
    \end{figure}
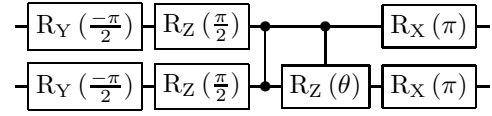

	Like its classical counterpart, a quantum circuit $U$ consists of a sequence (product) of \emph{quantum logic gates} $U_i$, i.e., $U= U_d\cdots U_1$. Here $d$ is the gate number of the circuit $U$. Each gate $U_i$ only non-trivially operates on one or two qubits. 
	We list commonly used 1-qubit  gates in Table~\ref{table:single}. Arbitrary 1-qubit gates can be decomposed into 1-qubit rotation gates $R_x(\theta)$, $R_y(\theta)$ and $R_z(\theta)$ with rotation parameter $\theta$. In addition, for any 1-qubit logic gate $U$, we can generate a 2-qubit logic gate --- controlled-$U$ (CU) gate, applying $U$ on the second (target) qubit 
	if and only if the first (control) qubit is $\ket{1}$. Specifically, the controlled-Z (CZ) gate (commonly used in superconducting quantum circuits) and the general controlled-$U$ gate are illustrated as follows. Fig.~\ref{fig:circuit_example} shows an example of a 2-qubit quantum circuit for the QAOA algorithm.
	{
		\begin{table}[]
			\centering
			\caption{\textsc{1-Qubit Gate}}
			\begin{tabular}{|c|c|c|c|}
				\hline
				H & $\frac{1}{\sqrt{2}}\begin{pmatrix}
					1 & 1 \\
					1 & -1
				\end{pmatrix}$ &
				X ($\sigma_x $) & $\begin{pmatrix}
					0 & 1 \\
					1 & 0
				\end{pmatrix}$ \\
				\hline
				Y ($\sigma_y $) & ${\left(\begin{matrix}
						0 & -i \\
						i & 0
					\end{matrix}\right)}$ &
				Z ($\sigma_z $) & $\left(\begin{matrix}
					1 & 0 \\
					0 & -1
				\end{matrix}\right)$ \\
				\hline
				T & ${\left(\begin{matrix}
						1 & 0 \\
						0 & e^{i \pi/4}
					\end{matrix}\right)}$ &
				$R_x(\theta)$ & ${\left(\begin{matrix}
						\cos \frac{\theta}{2} & -i \sin \frac{\theta}{2} \\
						-i \sin \frac{\theta}{2} & \cos \frac{\theta}{2}
					\end{matrix}\right)}$\\
				\hline
				$R_y(\theta)$ & ${\left(\begin{matrix}
						\cos \frac{\theta}{2} & -\sin \frac{\theta}{2} \\
						\sin \frac{\theta}{2} & \cos \frac{\theta}{2}
					\end{matrix}\right)}$ &
				$R_z(\theta)$ & ${\left(\begin{matrix}
						e^{-i \theta / 2} & 0 \\
						0 & e^{i \theta / 2}
					\end{matrix}\right)}$\\
				\hline
			\end{tabular}
			\label{table:single}
		\end{table}
	}
	\[
	\begin{array}{c}\begin{aligned}
			\Qcircuit @C=1.5em @R=1.5em {
				& \ctrl{1} & \qw\\
				&\gate{U} & \qw}
		\end{aligned}
  	=\left(\begin{matrix}1&0&0&0\\
			0&1&0&0\\
            0 & 0 & u_{00} & u_{01} \\
            0 & 0 & u_{10} & u_{11}
        \end{matrix}\right)\\
	\end{array}
	\begin{array}{c}\begin{aligned}
			\Qcircuit @C=1em @R=1em {
				& \ctrl{1} & \qw \\
				& \ctrl{-1} & \qw}
		\end{aligned}
		=\left(\begin{matrix}1&0&0&0\\
			0&1&0&0\\
			0&0&1&0\\
			0&0&0&-1
		\end{matrix}\right)\\
	\end{array}
	\]

	{\bf Quantum Noises:}  In the current NISQ era, quantum noise is unavoidable, so we have to consider the noisy effect in quantum computing. Consequently, the uncertainty will be brought such that quantum states become mixed instead of pure states.
 A \emph{mixed state} is considered as an ensemble $\{(p_k,|\psi_k\rangle\}$, meaning the quantum system is in state $|\psi_k\rangle$ with probability $p_k$.  Mathematically, it can be described by a $2^n\times 2^n$ \emph{density matrix} $\rho$ (Hermitian  positive semidefinite matrix with unit trace\footnote{$\rho$ has unit trace if $\tr(\rho) = 1$, where trace $\tr(\rho)$ of $\rho$ is defined as the summation of diagonal elements of $\rho$.}) on $\h$: 
	$\rho = \sum_{k} p_{k}\ket{\psi_{k}}\bra{\psi_{k}},$
	where  $\bra{\psi_{k}}$ is the conjugate  transpose of $\ket{\psi_{k}}$, i.e., $\bra{\psi_{k}}=\ket{\psi_{k}}^\dagger$. In this situation, a computational task starting at a mixed state $\rho$ is finished by a mapping $\e$:
	$\rho'=\e(\rho),$ where $\e$ is a quantum channel. A channel $\e$ (also called a \emph{super-operator}) admits a \emph{Kraus matrix form}~\cite{Nielsen2010}: there exists a finite set $\{E_k\}_{k\in\k}$ of matrices on $\h$ such that 
	\[\setlength\abovedisplayskip{1.5pt}
	\setlength\belowdisplayskip{1.5pt}
	\e(\rho)=\sum_{k\in\k}E_k\rho E_k^\dagger \quad \textrm{ with  } \sum_{k\in\k}E_k^\dagger E_k=I_n,\]
	where $\{E_k\}_{k\in\k}$ is called \emph{Kraus matrices} of $\e$. Specifically, quantum gate $U$ can be viewed as a unitary channel where $\e(\rho) = U \rho U^\dagger$. Briefly, $\e$ is represented as $\e=\{E_k\}_{k\in\k}$. Quantum noises can be represented as quantum channels. An example of quantum noise is the depolarizing noise, which is represented as $\{\sqrt{1-p}I, \sqrt{\frac{p}{3}}X, \sqrt{\frac{p}{3}}Y, \sqrt{\frac{p}{3}}Z\}$. Similar to noiseless quantum circuit $U$, noisy quantum circuit $\e$ also consists of a sequence (mapping composition) of  (noisy) gates $\{\e_{i}\}$, i.e., $\e=\e_d\circ\cdots \circ\e_1$, where each $\e_{i}$ is either a noiseless gate or a noise channel.

    The only way allowed by quantum mechanics to extract classical information is through a quantum measurement, which is mathematically modeled by a set $\{P_{i}\}_{i\in\o}$ of projection matrices on its state (Hilbert)  space $\h$ with $\o$ being the set of outputs and $\sum_{i
		\in\o} P_i=I_n$. This observing process is probabilistic: for a given quantum state $\rho$, a measurement outcome $i\in\o$ is obtained with probability $p_{i}=\tr(P_i\rho).$

	\section{Noisy Quantum Circuit Simulation}\label{sec:pre}
The simulation task for noiseless quantum circuit is to calculate the state vector representation of the output state of a circuit for a given input state. 
 For noisy quantum circuits, the simulation task targets to estimate the density matrix of the output state. Specifically, given a noisy quantum circuit $\e_{\N}$ and input state $\rho_0$, the goal of the noisy simulation task is to  estimate $\e_{\N}(\rho_0)$.  Every element of  $\e_{\N}(\rho_0)$ can be independently estimated by  $\bra{x}\e_{\N}(\rho_0)\ket{y}$, where  $\ket{x},\ket{y}$ are 
 pure states ranging from all computational basis of $\h$. Furthermore, we observe that  the value of $\bra{x}\e_{\N}(\rho_0)\ket{y}$ can be computed by the following way:
     \[
    \begin{aligned}
            \bra{x}|\e_{\N}(\rho_0)\ket{y} = & \frac{1}{4}((\bra{x}+\bra{y})|\e_{\N}(\rho_0)(\ket{x}+\ket{y})\\
                                    &- (\bra{x}-\bra{y})|\e_{\N}(\rho_0)(\ket{x}-\ket{y}) \\
                                    &- i(\bra{x}-i\bra{y})|\e_{\N}(\rho_0)(\ket{x}+i\ket{y}) \\
                                    &+ i(\bra{x}+i\bra{y})|\e_{\N}(\rho_0)(\ket{x}-i\ket{y})
    \end{aligned}
    \]
Therefore the key to simulating a noisy quantum circuit is as follows. 
	\begin{prob}[Noisy Simulation Task of Quantum Circuits]\label{prob:main}
		Given a noisy quantum circuit with $d$ gates $\e_{\N}=\e_d\circ\cdots\circ\e_1$~\cite{Nielsen2010}, an input state $\ket{\psi}$ and a state $\ket{ v }$, the \emph{noisy simulation of quantum cirucuit} $\e_{\N}$ on $\ket{\psi}$ is to estimate $\bra{ v }\e_{\N}(\ketbra{\psi}{\psi})\ket{ v }$ (with a high accuracy).
	\end{prob}

 This paper aims to solve the problem using tensor networks as the data structure for modeling quantum circuits and SVD for approximating the tensor representation of quantum noise. Tensor networks provide a graphical representation for quantum systems and their interactions. The tensor diagrams offer an intuitive way to understand and process the behavior of quantum circuits (see~\cite{biamonte2017tensor} for more details). In the following, we introduce a tensor network diagram for the noisy simulation task in Problem~\ref{prob:main}, which is the foundation for our approximation simulation algorithm. Furthermore, this diagram directly leads to a tensor network-based algorithm for exactly computing $\bra{ v }\e_{\N}(\ketbra{\psi}{\psi})\ket{ v }$.

{\bf Tensor Network Diagram for Noisy Simulation:} 	
To compute $\bra{ v }\e_{\N}(\ketbra{\psi}{\psi})\ket{ v }$, one can utilize the matrix representation of quantum super-operators~\cite[Chapter 2.2.2]{watrous2018theory}.
	\begin{equation*}
		\begin{aligned}
			\bra{ v }\e_{\N}(\ketbra{\psi}{\psi})\ket{ v } &= \tr( \ketbra{v}{v} \e_{\N}(\ketbra{\psi}{\psi}))\\
			&= \bra{\Omega}[\ketbra{ v ^*}{ v ^*}\otimes \e_{\N}(\ketbra{\psi}{\psi})]\ket{\Omega}\\
			&=\bra{\Omega}[\ketbra{ v ^*}{ v ^*}\otimes \e_d\circ\cdots\circ\e_1(\ketbra{\psi}{\psi})]\ket{\Omega} \\
			&= \bra{ v }\otimes\bra{ v ^*}(M_{\e_d}\cdots M_{\e_1}) \ket{\psi}\otimes\ket{\psi^*}\label{eq:serial}
		\end{aligned}
	\end{equation*}
	where $\ket{\psi^*}$ is the entry-wise conjugate of $\ket{\psi}$, $\ket{\Omega}$ is the (unnormalized) maximal entangled state, i.e., $\ket{\Omega}=\sum_{j}\ket{j}\otimes\ket{j}$ with $\{\ket{j}\}$ being an orthonormal basis of Hilbert space $\h$, and $M_\e=\sum_{k}E_k\otimes E_k^*$ is called the \emph{matrix representation} of $\e$, where $\e(\rho) = \sum\limits_k E_k\rho E_{k}^\dagger$ with Kraus operators $\{E_k\}_{k\in\k}$. The tensor diagram can visualize this representation.
 
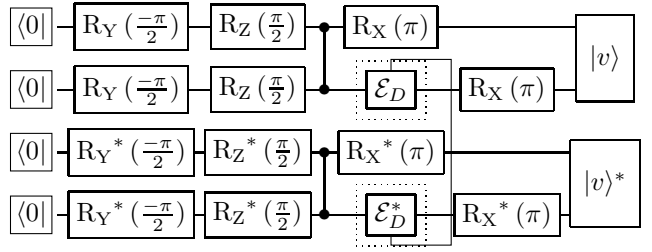
\begin{figure}
        \begin{align*}
        \centering
        \begin{array}{c}
        \Qcircuit @C=0.4em @R=0.6em @!R {\\
             \lstick{\tikzdiagram{\node[fill=white,tensor,overlay,anchor=base] {$\bra{0}$};}} & \gate{\mathrm{R_Y}\,(\mathrm{\frac{-\pi}{2}})} & \gate{\mathrm{R_Z}\,(\mathrm{\frac{\pi}{2}})} & \ctrl{1} & \gate{\mathrm{R_X}\,(\mathrm{\pi})} & \qw & \multigate{1}{\ket{v}}\\
             \lstick{\tikzdiagram{\node[fill=white,tensor,overlay,anchor=base] {$\bra{0}$};}} & \gate{\mathrm{R_Y}\,(\mathrm{\frac{-\pi}{2}})} & \gate{\mathrm{R_Z}\,(\mathrm{\frac{\pi}{2}})} & \control\qw & \gate{\tikz{\node[inner sep=0] (a){$\e_D$};}} & \gate{\mathrm{R_X}\,(\mathrm{\pi})} & \ghost{\ket{v}}\\
             \lstick{\tikzdiagram{\node[fill=white,tensor,overlay,anchor=base] {$\bra{0}$};}} & \gate{\mathrm{R_Y}^*\,(\mathrm{\frac{-\pi}{2}})} & \gate{\mathrm{R_Z}^*\,(\mathrm{\frac{\pi}{2}})} & \ctrl{1} & \gate{\mathrm{R_X}^*\,(\mathrm{\pi})} & \qw & \multigate{1}{\ket{v}^*}\\
             \lstick{\tikzdiagram{\node[fill=white,tensor,overlay,anchor=base] {$\bra{0}$};}}  & \gate{\mathrm{R_Y}^*\,(\mathrm{\frac{-\pi}{2}})} & \gate{\mathrm{R_Z}^*\,(\mathrm{\frac{\pi}{2}})} & \control\qw & \gate{\tikz{\node[inner sep=0] (b){$\e_D^*$};}} & \gate{\mathrm{R_X}^*\,(\mathrm{\pi})} & \ghost{\ket{v}^*}
            \gategroup{3}{5}{3}{5}{.75em}{.}
            \gategroup{5}{5}{5}{5}{.75em}{.}
             }
        \end{array}
        \end{align*}
        \tikz[overlay]{\draw[shorten > = 3pt, shorten < = 3pt] (a.north) |- +(0.8,0.25) |- ($(b)+(0.5,-0.4)$) -| (b.south);}
        \caption{Tensor network diagram of the 2-qubit QAOA circuit in Fig.~\ref{fig:circuit_example} with input state $\ket{0}$.}
        \label{fig:circuit_double_example}
    \end{figure}

	We note that $M_{\e}=\sum_{k\in\k}E_{k}\otimes E_{k}^*$ represents the noises in the newly obtained tensor network. In particular, for a unitary super-operator $\u(\rho)=U\rho U^\dagger$ with unitary operator $U$, $M_\u$ is depicted as the right tensor network in the following. 
	\begin{align*}
    \tiny
		M_{\e}=\sum_{k\in\k}E_{k}\otimes E_{k}^*=\begin{aligned}
			\Qcircuit @C=1em @R=1em{
				&\gate{E_k} & \qw\\
				&\gate{E_k*} \qwx &\qw}\end{aligned} \qquad
		M_{\u}=U\otimes U^*=\begin{aligned}
			\Qcircuit @C=1em @R=1em{
				&\gate{U} & \qw\\
				&\gate{U^*} &\qw}\end{aligned}
	\end{align*}
	Based on these observations, we get a serial connection of the two tensor networks representing $n$-qubit circuits $\e_{\N}$. Subsequently, we derived an accuracy algorithm to compute $\bra{ v }\otimes\bra{ v ^*}(M_{\e_d}\cdots M_{\e_1}) \ket{\psi}\otimes\ket{\psi^*}$ by contracting a tensor network with double size ($2n$ qubits). An illustration of the tensor network diagram of the two-qubit noisy QAOA circuit of Fig.~\ref{fig:circuit_example} is shown in Fig.~\ref{fig:circuit_double_example}.
 
\section{Approximation Algorithm For Noisy Quantum Circuit Simulation}\label{sec:algorithm}
    We are ready to present our approximation noisy simulation algorithm based on the new tensor network diagram.
    
	{\bf Approximation Noisy Simulation Algorithm:}
	As we can see, the noises make simulating a quantum circuit much harder. To handle the difficulty of simulating a circuit with a large number of noises, we introduce an approximation noisy circuit simulation method based on the matrix representation and the SVD (Singular Value Decomposition), which can balance the accuracy and efficiency of the simulation. The insight of our method starts with an observation that most noises occurring in physical quantum circuits are close to the identity operators (matrices). Take the depolarizing noise as an example, which is defined by
	\[
	\mathcal{E}(\rho)=(1-p) \rho+\frac{p}{3}(X \rho X+Y \rho Y+Z \rho Z).
	\]
    Probability $p$ can be regarded as a metric of the noisy effectiveness of the channel. When $p$ is small, the noise is almost identical to the identity channel. Therefore, a straightforward method is to approximately represent the depolarizing channel by $(1-p)I$.
The current physical implementation of practical quantum algorithms requires that the effectiveness of noise is insignificant.
For general noise $\e$, we define $\|M_\e - I\|$  as the \emph{noise rate} of $\e$, where $\|\cdot\|$ is the 2-norm of matrix. For example, the depolarizing noise with parameter $p$ has a noise rate $2p$. Next, we will describe our approximation algorithm and explain why it is efficient when the noise rate is small.

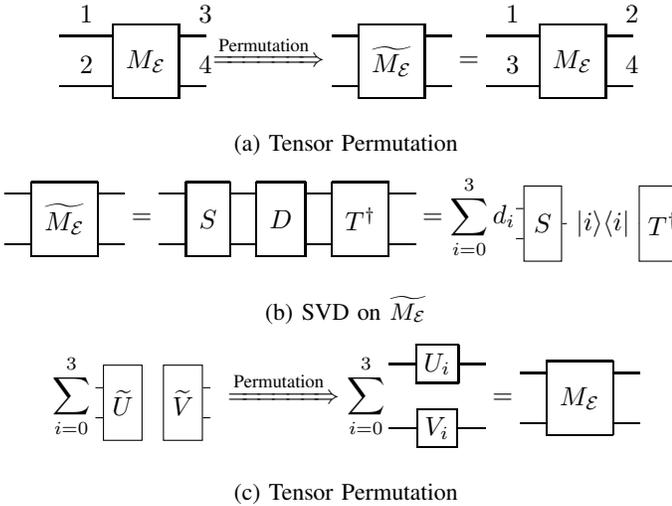
\begin{figure}[!ht]
		\begin{subfigure}{.5\textwidth}
			\begin{align*}
				&\begin{aligned}
					\Qcircuit @C=1em @R=1em{
						& \ustick{1} \qw  & \multigate{1}{M_\e}   & \ustick{3} \qw\\
						& \ustick{2} \qw  & \ghost{M_\e}          & \ustick{4} \qw\\}
				\end{aligned}
				\xRightarrow{\text{Permutation}} \begin{aligned}
					\Qcircuit @C=1em @R=1em{
						& \multigate{1}{\widetilde{M_\e}} & \qw \\
						& \ghost{\widetilde{M_\e}} & \qw \\}
				\end{aligned}
				= \begin{aligned}
					\Qcircuit @C=1em @R=1em{
						& \ustick{1} \qw &\multigate{1}{M_\e} & \ustick{2} \qw\\
						& \ustick{3} \qw &\ghost{M_\e} & \ustick{4} \qw}
				\end{aligned}
			\end{align*}
			\caption{Tensor Permutation}
		\end{subfigure}
		\begin{subfigure}{.5\textwidth}
			\begin{align*}
				&\begin{aligned}
					\Qcircuit @C=1em @R=1em{
						& \multigate{1}{\widetilde{M_\e}} & \qw \\
						& \ghost{\widetilde{M_\e}} & \qw \\}
				\end{aligned}
				= \begin{aligned}
					\Qcircuit @C=1em @R=1em{
						& \multigate{1}{S} & \multigate{1}{D} & \multigate{1}{T^\dagger} & \qw\\
						& \ghost{S} & \ghost{D} & \ghost{T^\dagger} & \qw\\
					}
				\end{aligned}
				=  \sum\limits_{i=0}^3 d_i \raisebox{-0.5\height}{%
					\begin{tikzpicture}
						\node[draw,rectangle,minimum width=0.5cm, minimum height = 1cm] (u) at (0,1) {$S$};
						\node[draw,rectangle,minimum width=0.5cm, minimum height = 1cm] (v) at (1.6,1) {$T^\dagger$};
						\node (D) at (0.8,1) {$\ketbra{i}{i}$};
						\draw (u) -- (D);
						\draw (D) -- (v);
						\draw (u.west) ++ (0, 0.2) -- ++(-0.1,0);
						\draw (u.west) ++ (0, -0.2) -- ++(-0.1,0);
						\draw (v.east) ++ (0, 0.2) -- ++(0.1,0);
						\draw (v.east) ++ (0, -0.2) -- ++(0.1,0);
					\end{tikzpicture}
				}
			\end{align*}
			\caption{SVD on $\widetilde{M_\e}$}
		\end{subfigure}
		\begin{subfigure}{.5\textwidth}
			\begin{align*}
				\sum\limits_{i=0}^3\raisebox{-0.5\height}{%
					\begin{tikzpicture}
						\node[draw,rectangle,minimum width=0.5cm, minimum height = 1cm] (u) at (0,1) {$\widetilde{U}$};
						\node[draw,rectangle,minimum width=0.5cm, minimum height = 1cm] (v) at (0.8,1) {$\widetilde{V}$};
						\draw (u.west) ++ (0, 0.2) -- ++(-0.1,0);
						\draw (u.west) ++ (0, -0.2) -- ++(-0.1,0);
						\draw (v.east) ++ (0, 0.2) -- ++(0.1,0);
						\draw (v.east) ++ (0, -0.2) -- ++(0.1,0);
					\end{tikzpicture}
				}
				\xRightarrow{\text{Permutation}} 
				\sum\limits_{i=0}^3\begin{aligned}
					\Qcircuit @C=1em @R=1em{
						& \gate{U_i} & \qw \\
						& \gate{V_i} & \qw \\}
				\end{aligned}
				=\begin{aligned}
					\Qcircuit @C=1em @R=1em{
						& \multigate{1}{M_\e} & \qw \\
						& \ghost{M_\e} & \qw \\}
				\end{aligned}
			\end{align*}
			\caption{Tensor Permutation }	
		\end{subfigure}
		\caption{Decomposition of $M_\e$}
		\label{fig:svd}
	\end{figure}

To handle the noises, we perform SVD on the matrix representation of each noise as illustrated in Fig.~\ref{fig:svd}.

Suppose we treat $M_{\e}$ as a tensor with rank 4, and the edges are indexed as in Fig.~\ref{fig:svd} (a). $M_{\e}$ is the matrix with edges $1, 2$ being the row and $3, 4$ being the column. We introduce the tensor permutation operator where $\tilde{M_{\e}}$ is the matrix with $1,3$ being the row and $2, 4$ being the column. For example, the identity operator $I$ on two qubits  and its tensor permutation $\tilde{I}$ are:
	\[
	\begin{array}{c}
		I = \left(\begin{matrix}
			1 & 0 & 0 & 0 \\
			0 & 1 & 0 & 0 \\
			0 & 0 & 1 & 0 \\
			0 & 0 & 0 & 1
		\end{matrix}\right)
	\end{array}
	\begin{array}{c}
		\tilde{I} = \left(\begin{matrix}
			1 & 0 & 0 & 1 \\
			0 & 0 & 0 & 0 \\
			0 & 0 & 0 & 0 \\
			1 & 0 & 0 & 1
		\end{matrix}\right)
	\end{array}
	\]
    Performing SVD on $\tilde{M_{\e}}$ we have $\tilde{M_{\e}} = S D T^\dagger$, where $D = \sum\limits_{i=0}^3 d_i \ketbra{i}{i}$ and $d_0$ is the most significant singular value (see Fig.~\ref{fig:svd}(b)). Let $\tilde{U_i} = d_i S\ket{i}$ and $\tilde{V_i} = T\ket{i}$ as in the left of Fig.~\ref{fig:svd} (c), we get $\tilde{M_{\e}} = \sum\limits_{i=0}^3 \tilde{U_i} \tilde{V_i}^\dagger$. Then use tensor permutation again on $\tilde{U_i}$ and $\tilde{V_i}$ (here $\tilde{U_i}$ is treated as a rank 4 tensor with index 3,4 being 0-dimension, and $\tilde{U_i}$ is treated as a rank 4 tensor with index 1,2 being 0-dimension), we get $M_{\e} = \sum\limits_{i = 0, 1, 2, 3} U_i \otimes V_i$, as illustrated in Fig.~\ref{fig:svd} (c).
	
	Next, we will show that when the noise rate of $\e$ is small, $U_0 \otimes V_0$ is a good approximation of $M_\e$.
	
	\begin{lem}
		Suppose $A$ and $B$ are all $4 \times 4$ matrices, and $\tilde{A}$ and $\tilde{B}$ are the tensor permutation of $A$ and $B$ as defined before. If $\|A - B\| < \delta$, then we have $\|\tilde{A} - \tilde{B}\| < 2 \delta$.  
	\end{lem}
	\begin{proof}
		We use $\|\cdot\|_{\mathrm{F}}$ to donate the Frobenius norm of matrices. Specifically, for $4 \times 4$ matrices, it can be shown that $\|A\| \leq\|A\|_F \leq 2\|A\|$.
		Note that $A$ and $\tilde{A}$ have the same Frobenius norm since the tensor permutation operator only rearranges the positions of elements.
		Therefore we have $\|\tilde{A} - \tilde{B}\| \leq \|\tilde{A} - \tilde{B}\|_F = \|A - B\|_F \leq 2\|A-B\| <2\delta$.
	\end{proof}
	\begin{lem}
		Suppose $\|M_\e - I\| < \delta$, then we have $\|M_\e - U_0 \otimes V_0 \| < 4\delta$
	\end{lem}
	\begin{proof}
		By Lemma 1. we have $\|\tilde{M_\e} - \tilde{I}\| < 2 \delta$. Suppose we perform SVD on $\tilde{M_{\e}}$ we have $\tilde{M_{\e}} = S D T^\dagger$. And let $D_0 = d_0 \ketbra{0}{0}$, we have 
		\[
		\|\tilde{M_\e} - S D_0 T^\dagger\| = \min_{rank(A) = 1} \| \tilde{M_\e} - A \| \leq \|\tilde{M_\e} - \tilde{I}\| < 2\delta.
		\]
        The first equation comes from the Eckart-Young-Mirsky theorem~\cite{eckart1936approximation}. The last inequality holds because $\tilde{I}$ has rank 1.
		And note that $\tilde{\tilde{M}}_\e = M_\e$ and $\widetilde{S D_0 T^\dagger} = U_0 \otimes V_0$, by Lemma 1, we have 
		\[
		\|M_\e -  U_0 \otimes V_0\| < 4\delta. \qedhere
		\]
    \end{proof}
	We are ready to introduce our approximation algorithm for the noisy quantum circuit simulation task. Suppose $\e_{\N}=\e_d\circ\cdots\circ\e_1$ is a quantum circuit with $N$ noises, where the noise channels have index $\{i_1, ..., i_N\}$. After applying SVD on each matrix representation of noises we get $M_{\e_{i_s}} = U_0^s \otimes V_0^s + U_1^s \otimes V_1^s + U_2^s \otimes V_2^s + U_3^s \otimes V_3^s, s \in \{1,...,N\}$, where $M_{\e_{i_s}}$ is closed to $U_0^s \otimes V_0^s$. The idea of our algorithm is to calculate the simulation result by using $U_0^s \otimes V_0^s$ to substitute the noises as a prior choice. Let $M_{\e_{i_s}}' = U_0^s \otimes V_0^s$ and $\overline{M_{\e_{i_s}}} = U_1^s \otimes V_1^s + U_2^s \otimes V_2^s + U_3^s \otimes V_3^s$. Thus $M_{\e_{i_s}} = M_{\e_{i_s}}' + \overline{M_{\e_{i_s}}}$ and $\|\overline{M_{\e_{i_s}}}\| < 4p$ by Lemma 2. Let $T_u$ be the sum of tensor networks obtained by substituting all but $u$ noises to $M_{\e_{i_s}}'$ and $u$ noises to one of $U_i^s \otimes V_i^s, i = 1, 2, 3$, i.e., $T_u = \sum\limits_{1 \leq p_1 < \cdots < p_u \leq N} \prod\limits_{r = 1}^{u}\overline{M_{\e_{i_{p_r}}}} \prod\limits_{t \notin \{p_1,...p_u\}} M_{\e_{i_t}}'$. We call $A(l) = \sum\limits_{u=0}^l T_u$ the \emph{l-level approximation} of $\e_{\N}$. By increasing the approximation level $l$ we get a better approximation for $\e_{\N}$, when $l=N$, $A(l) = M_{\e_{\N}}$ exactly. Our algorithm is to calculate $A(l)$ for a given $l$. Details of the algorithm are shown in Algorithm~\ref{Algorithm:Approximate}.
	
	\begin{algorithm}[htbp]
		\caption{ApproximationNoisySimulation($\e_{\N},\ket{\psi},\ket{ v }, l$)}
		\label{Algorithm:Approximate}
		{\small{
				\begin{algorithmic}[1]
					\Require A noisy quantum circuit $\e_{\N}=\e_d\circ\cdots\circ\e_1$ with $N$ noises, where the noise channels have index $\{i_1, ..., i_N\}$ and Kraus matrices $\e_{i_s}=\{E_{sk}\}_{k\in\k_s}$ for $1 \leq s \leq N$, a test state $\ket{\psi}$, an expected output $\ket{ v }$ and a level for the approximation $l$.
					\Ensure $A(l)$, the $l$-level approximation of $\bra{ v }\e_{\N}(\ketbra{\psi}{\psi})\ket{ v }$
					\State Result = 0
					\State Generate the tensor network as in the accuracy algorithm.
					\State Perform tensor permutation and SVD operation to the matrix representation of all noises $M_{\e_{i_s}}$, obtaining $U_k^s$ and $V_k^s$, where $M_{\e_{i_s}} = U_0^s \otimes V_0^s + U_1^s \otimes V_1^s + U_2^s \otimes V_2^s + U_3^s \otimes V_3^s, s \in \{1,...,N\}$.
					\ForAll {$0 \leq k \leq l$}
					\State Calculate approximation of level $k$:
                    \State $T_k := 0$
					\ForAll{$1 \leq p_1 < \cdots < p_k \leq N$}
					\State Substitute $M_{\e_{i_{p_j}}}$ with one of $U_1^{p_j} \otimes V_1^{p_j}, U_2^{p_j} \otimes V_2^{p_j}, U_3^{p_j} \otimes V_3^{p_j}$ for $j \in \{1,...k\}$. For $s \notin \{p_1, p_2, ..., p_k\}$, substitute $M_{\e_{i_s}}$ to $U_0^s \otimes V_0^s$.
                    \State After substitution, the original (double-sized) tensor network is split into two tensor networks, contract both tensor networks, and multiply the result, donated as $X$.
					\State $T_k$ += X.
					\EndFor
                    \State Result += $T_k$.
                    \EndFor
					\State\Return Result
				\end{algorithmic} 
		}}
	\end{algorithm}

	\begin{thm}\label{thm: precision}
		Given a noisy quantum circuits $\e_{\N}$ with $d$ gates and $N$ noises with all noise rates being  less than $p$, i.e., $\|M_{\e_{i_s}} - I\| < p$ for $s \in \{1,...,N\}, i_s \in \{1, ..., d\}$. For any input and output state $\ket{\psi}$ and $\ket{v}$, donate the fidelity as $F = \bra{ v }\e_{\N}(\ketbra{\psi}{\psi})\ket{ v }$, we have $|F - A(l)| < (1+8p)^N - \sum\limits_{i = 0}^{l} \tbinom{N}{i}(4p)^i(1+4p)^{(N-i)}$, 
  and the number of tensor network contractions is $2\sum\limits_{i = 0}^{l} \tbinom{N}{i}3^i$.
	\end{thm}
	\begin{proof}
	\begin{equation*}
	\begin{aligned}
		|F - A(l)|
  &=\|M_{\e_N} \cdots M_{\e_1} - \sum\limits_{i = 0}^{l} T_i\|\\
  &= \|\sum\limits_{i = l+1}^{N} T_i\|\\
  &\leq \sum\limits_{i = l+1}^{N} \|T_i\|\\
		&\leq \sum\limits_{i = l+1}^{N} \tbinom{N}{i}(4p)^i(1+4p)^{(N-i)}\\
		&= (1+8p)^N - \sum\limits_{i = 0}^{l} \tbinom{N}{i}(4p)^i(1+4p)^{(N-i)}\\
	\end{aligned}
	\end{equation*}
	Therefore
    \begin{align*}
            &\bra{ v }\otimes\bra{ v ^*}(M_{\e_{\N}} - M_{\e_{\N}}') \ket{\psi}\otimes\ket{\psi^*}\\
            \leq &\|\ket{ v }\otimes\bra{ v ^*}\| \|(M_{\e_{\N}} - M_{\e_{\N}}') \ket{\psi}\otimes\ket{\psi^*}\| \leq \|M_{\e_{\N}} - M_{\e_{\N}}'\|&& \\
            = &(1+8p)^N - \sum\limits_{i = 0}^{l} \tbinom{N}{i}(4p)^i(1+4p)^{(N-i)} \qedhere
    \end{align*}
\end{proof}
	Specifically, when using the one-level approximation where the number of noises  is $N$ and the noise rate is $p$, Theorem~\ref{thm: precision} asserts that our method has accuracy $(1+8p)^N - 1 - 4Np(1+4p)^{N-1}$. With the assumption that $p \leq \frac{1}{8N}$, we have 
    \[
        \begin{aligned}
            &(1+8p)^N - 1 - 4Np(1+4p)^{N-1}\\
            =&\sum\limits_{k=2}^N \tbinom{N}{k}(4p)^k(1+4p)^{N-k}\\
            \leq& (1+\frac{1}{2N})^N \sum\limits_{k=2}^N (4Np)^k\\
            \leq& \sqrt{e} (4Np)^2 \frac{1-(4Np)^{N-1}}{1-4Np}\\
            \leq& 32\sqrt{e}N^2p^2
        \end{aligned}
    \]
    As a comparison, the quantum trajectories method~\cite{isakov2021simulations} needs $r$ samples (tensor network contraction) to obtain a result within accuracy $O(1/\sqrt{r})$ under a constant possibility. For the quantum trajectories method to achieve such accuracy (under a constant probability), $N^2p^2 = \frac{C}{\sqrt{r}}$. Therefore the sample number $r = \frac{C^2}{N^4p^4}$, where $C$ is a constant number. Note that our level-1 approximation needs $O(N)$ samples, and therefore our approximation method will need less sampling number than the Monte-Carlo method when $p = O(N^{-\frac{5}{4}})$.
	
	\section{Experiments}
	In this section, we demonstrate the utility and effectiveness of our approximation algorithm by simulating various quantum circuits with realistic noise models. In particular, we compare our algorithm with state-of-the-art
accurate and approximate methods for the same task, and further numerically analyze our algorithm in terms of accuracy, approximation levels, and noise rate. The code of our implementation can be accessed at Github repository \footnote{\href{https://github.com/Veri-Q/VeriQSim}{https://github.com/Veri-Q/VeriQSim}}.
 
	{\bf Runtime Environment:} All our experiments are carried out on a server with Intel Xeon Platinum 8153 @ $2.00 \mathrm{GHz} \times 256$ Cores, 2048 GB Memory. The machine runs CentOS 7.7.1908. We use the Google TensorNetwork Python package~\cite{roberts2019tensornetwork} for the tensor network computation. 
	
	{\bf Benchmark Circuits and Fault Models:}
Our benchmark circuits consist of three types of quantum circuits: Quantum Approximate Optimization Algorithm (QAOA), Hartree-Fock Variational Quantum Eigensolver (VQE), and random quantum circuits exhibiting quantum supremacy from Google. Google has experimentally run all three types of quantum circuits on their quantum processors. In our experiments, the benchmark circuits are taken from ReCirq~\cite{quantum_ai_team_and_collaborators_2020_4091470}, an open-source library for Cirq and Google's Quantum Computing Service. The QAOA and VQE circuits are named \textit{\textbf{qaoa\_N}} and \textit{\textbf{hf\_N}} respectively, where $N$ denotes the number of qubits. The random quantum supremacy circuits with $R\times C$ qubits and depth $D$ are named \textit{\textbf{inst\_$R\times C$\_$D$}}. 

We use a realistic decoherence noise model of superconducting quantum circuits to model faults~\cite{fault_model}. Each decoherence noise is appended after a randomly chosen gate in the circuit. This emulates the types of errors seen on actual quantum hardware.

The benchmarks and noise models provide a real-world testbed for our techniques. By evaluating our approximation algorithm on these practical circuits and fault models, we can obtain a practical performance assessment of the algorithm for accurately and efficiently simulating the real implementation of quantum algorithms on NISQ devices.

    {\bf Baselines:} For evaluating the performance of our approximation algorithm, the baselines include the state-of-the-art accurate and approximate methods for simulating (noisy) quantum circuits. 
    
    The accurate methods include three types of noisy simulation algorithms, based on different data structures:
    \begin{itemize}
        \item \emph{MM-based method: matrix multiplication-based method} has been widely used in the field of quantum simulation. In this case, input quantum states, gates and noises in simulated circuits are all treated as matrices, and the simulation is executed by matrix multiplications.  
        \item \emph{TDD-based method: tensor decision diagram-based method} uses decision diagrams style method for representing quantum states, gates and noises as tensors~\cite{hongxinTDD}. Subsequently, it is more efficient in equivalence checking of quantum circuits~\cite{hong2021approximate}.
        \item \emph{TN-based method: tensor network-based method} uses tensors to represent quantum states, gates and noises such that quantum circuits can be converted into a network of tensors. Then the noisy simulation is finished by contracting tensors in the network. The efficiency is highly dependent on the contraction order. In the following experiment, it is shown that the TN-based method is more efficient than the MM-based and TDD-based methods. 
    \end{itemize}
    All the above methods will provide accurate simulation for noisy quantum circuits.
    
    The approximate method is the quantum trajectories method~\cite{isakov2021simulations}. It relies on a Monte Carlo approach, sampling the Kraus operators of a quantum noise based on probabilities. Consequently, it is a stochastic method and achieves a given accuracy probabilistically, while our approximation method is deterministic.
   
 {\bf Comparable Experiments:}
  We evaluate the efficiency of the aforementioned methods and our approximate techniques on benchmark circuits with realistic noises.  The memory out (MO) limit is capped at 2048 GB.
\begin{table*}[tbp]
\centering
\caption{\textsc{Our Algorithm vs. Accurate Methods}}
\begin{tabular}{|c|c|c|c|c|cccccc|}
\hline
\multirow{3}{*}{Type}      & \multirow{3}{*}{Circuit} & \multirow{3}{*}{Qubits} & \multirow{3}{*}{Gates} & \multirow{3}{*}{Depth} & \multicolumn{6}{c|}{Time(s)}                                                                                                                                                   \\ \cline{6-11} 
                           &                          &                         &                        &                        & \multicolumn{4}{c|}{\#Noise = 2}                                                                                                & \multicolumn{2}{c|}{\#Noise = 20}            \\ \cline{6-11} 
                           &                          &                         &                        &                        & \multicolumn{1}{c|}{MM}     & \multicolumn{1}{c|}{TDD}    & \multicolumn{1}{c|}{TN}          & \multicolumn{1}{c|}{Ours} & \multicolumn{1}{c|}{TN}      & Ours \\ \hline
\multirow{4}{*}{HF-VQE}    & hf\_6                    & 6                       & 155                    & 72                     & \multicolumn{1}{c|}{0.17}   & \multicolumn{1}{c|}{1.2}    & \multicolumn{1}{c|}{0.095}       & \multicolumn{1}{c|}{0.81}        & \multicolumn{1}{c|}{0.093}   & 15.74         \\ \cline{2-11} 
                           & hf\_8                    & 8                       & 308                    & 124                    & \multicolumn{1}{c|}{0.24}   & \multicolumn{1}{c|}{3.65}   & \multicolumn{1}{c|}{0.33}        & \multicolumn{1}{c|}{1.52}        & \multicolumn{1}{c|}{0.31}    & 28.63         \\ \cline{2-11} 
                           & hf\_10                   & 10                      & 461                    & 142                    & \multicolumn{1}{c|}{26.91}  & \multicolumn{1}{c|}{7.59}   & \multicolumn{1}{c|}{0.37}        & \multicolumn{1}{c|}{2.81}        & \multicolumn{1}{c|}{0.39}    & 40.11         \\ \cline{2-11} 
                           & hf\_12                   & 12                      & 690                    & 194                    & \multicolumn{1}{c|}{206.37} & \multicolumn{1}{c|}{18.81}  & \multicolumn{1}{c|}{0.99}        & \multicolumn{1}{c|}{3.56}        & \multicolumn{1}{c|}{0.98}    & 64.28         \\ \hline
\multirow{3}{*}{QAOA}      & qaoa\_64                 & 64                      & 1696                   & 42                     & \multicolumn{1}{c|}{MO}      & \multicolumn{1}{c|}{58.33}  & \multicolumn{1}{c|}{3.51}        & \multicolumn{1}{c|}{15.96}       & \multicolumn{1}{c|}{47.38}   & 182.32        \\ \cline{2-11} 
                           & qaoa\_121                & 121                     & 3322                   & 42                     & \multicolumn{1}{c|}{MO}      & \multicolumn{1}{c|}{225.76} & \multicolumn{1}{c|}{9.77}        & \multicolumn{1}{c|}{49.96}       & \multicolumn{1}{c|}{1131.61} & 408.05        \\ \cline{2-11} 
                           & qaoa\_225                & 225                     & 6330                   & 42                     & \multicolumn{1}{c|}{MO}      & \multicolumn{1}{c|}{MO}     & \multicolumn{1}{c|}{925.87}      & \multicolumn{1}{c|}{2052.45}     & \multicolumn{1}{c|}{MO}      & 6403.95       \\ \hline
\multirow{9}{*}{Supermacy} & inst\_4x4\_10            & 16                      & 115                    & 11                     & \multicolumn{1}{c|}{MO}      & \multicolumn{1}{c|}{0.88}   & \multicolumn{1}{c|}{0.07}        & \multicolumn{1}{c|}{0.75}        & \multicolumn{1}{c|}{0.08}    & 12.41         \\ \cline{2-11} 
                           & inst\_4x4\_40            & 16                      & 394                    & 41                     & \multicolumn{1}{c|}{MO}      & \multicolumn{1}{c|}{TO$_{1}$}     & \multicolumn{1}{c|}{4.34}        & \multicolumn{1}{c|}{26.04}       & \multicolumn{1}{c|}{MO}      & 63.42         \\ \cline{2-11} 
                           & inst\_4x4\_80            & 16                      & 764                    & 81                     & \multicolumn{1}{c|}{MO}      & \multicolumn{1}{c|}{TO$_{1}$}     & \multicolumn{1}{c|}{11.26}       & \multicolumn{1}{c|}{39.69}       & \multicolumn{1}{c|}{MO}      & 145.44        \\ \cline{2-11} 
                           & inst\_4x5\_10            & 20                      & 145                    & 11                     & \multicolumn{1}{c|}{MO}      & \multicolumn{1}{c|}{1.52}   & \multicolumn{1}{c|}{0.10}        & \multicolumn{1}{c|}{1.42}        & \multicolumn{1}{c|}{0.1}     & 11.30         \\ \cline{2-11} 
                           & inst\_4x5\_20            & 20                      & 261                    & 21                     & \multicolumn{1}{c|}{MO}      & \multicolumn{1}{c|}{TO$_{1}$}     & \multicolumn{1}{c|}{0.30}        & \multicolumn{1}{c|}{2.84}        & \multicolumn{1}{c|}{3725.2}  & 21.09         \\ \cline{2-11} 
                           & inst\_4x5\_80            & 20                      & 959                    & 81                     & \multicolumn{1}{c|}{MO}      & \multicolumn{1}{c|}{TO$_{1}$}     & \multicolumn{1}{c|}{TO$_{1}$} & \multicolumn{1}{c|}{TO$_{1}$}          & \multicolumn{1}{c|}{MO}      & 26936         \\ \cline{2-11} 
                           & inst\_6x6\_10            & 36                      & 264                    & 11                     & \multicolumn{1}{c|}{MO}      & \multicolumn{1}{c|}{0.77}   & \multicolumn{1}{c|}{0.22}        & \multicolumn{1}{c|}{5.71}        & \multicolumn{1}{c|}{0.26}    & 356.82        \\ \cline{2-11} 
                           & inst\_6x6\_20            & 36                      & 483                    & 21                     & \multicolumn{1}{c|}{MO}      & \multicolumn{1}{c|}{21.41}  & \multicolumn{1}{c|}{4.85}        & \multicolumn{1}{c|}{103.80}      & \multicolumn{1}{c|}{MO}      & 428.74        \\ \cline{2-11} 
                           & inst\_7x7\_10            & 49                      & 364                    & 11                     & \multicolumn{1}{c|}{MO}      & \multicolumn{1}{c|}{1.66}   & \multicolumn{1}{c|}{0.45}        & \multicolumn{1}{c|}{2.22}        & \multicolumn{1}{c|}{0.75}    & 21.95         \\ \hline
\end{tabular}
\label{table:vqe}
\end{table*}

\emph{Our Algorithm vs. Accurate Methods:}
First, we compare the three accurate methods on benchmark circuits with the number of noise being 2, and $\ket{\psi}$ and $\ket{ v }$ are all chosen to be $\ket{0}\otimes\cdots\otimes\ket{0}$. The result is shown in the columns of \#Noise = 2 in  Table~\ref{table:vqe}, where the timeout (TO$_{1}$) thresholds are set at 3,600 seconds. As we can see from the table, 
 the TN-based method outperforms the other methods, including our approximation algorithm in all three benchmarks. Thus, the TN-based method works very well for simulating noisy quantum circuits, so there is no need to apply our approximation algorithm. For complicated quantum circuits with a larger number of noise, the TN-based method may fail, but our approximation algorithm works well. 
 
 To see this, we reset the noise number to 20, then we get the result in the columns of \#Noise = 20 in Table~\ref{table:vqe}, where the timeout (TO$_{2}$) threshold is set to be 36,000 seconds. From the result, \emph{our approximation algorithm is more efficient than the TN-based method in the complex quantum circuits (with a larger number of qubits and depths)}, for example, see the rows of $qaoa\_121,qaoa\_225, inst\_4\times 5\_80$ and $inst\_6\times 6\_20$. 
 
\begin{figure}[htbp]
 	\centering
        \scalebox{.8}{
		\begin{tikzpicture}[scale = 0.8]
			\begin{axis}[
				xlabel=Noise Number,
				ylabel=Time(s),
                legend pos = north east
				]
				\addplot[smooth,mark=*,blue] plot coordinates {
					(2,6.97)
					(4,7.06)
					(6,7.15)
					(8, 7.17)
					(10, 14.61)
					(12, 24.74)
					(14, 41.98)
					(16, 48.42)
                    (18, 146.33)
                    (20, 383.47)
                    (22, 743.69)
                    (24, 1127.73)
					(26, 983.69)
                    (28, 1457.76)
                    (30, 3669.61)
				};
				\addlegendentry{TN-based}
				\addplot[smooth,mark=x,red] plot coordinates {
                    (0, 4.686998128890991)
                    (2, 33.0309317111969)
                    (4, 62.039101123809814)
                    (6, 90.69518446922302)
                    (8, 119.43770623207092)
                    (10, 147.86366391181946)
                    (12, 175.68643736839294)
                    (14, 205.7167830467224)
                    (16, 233.2696704864502)
                    (18, 263.63577604293823)
                    (20, 291.5200433731079)
                    (22, 322.27098202705383)
                    (24, 347.222158908844)
                    (26, 375.38722229003906)
                    (28, 403.95211839675903)
                    (30, 435.1446626186371)
                    (32, 461.0113615989685)
                    (34, 493.12047052383423)
                    (36, 537.988543510437)
                    (38, 550.2323930263519)
                    (40, 589.8075199127197)                    
                    (42, 612.3398413658142)
                    (44, 643.3014228343964)
                    (46, 669.5444548130035)
                    (48, 700.7802011966705)
                    (50, 728.9921412467957)
                    (52, 759.2806141376495)
                    (54, 790.9490053653717)
                    (56, 822.5740852355957)
                    (58, 846.6838459968567)
                    (60, 881.5071609020233)
                    (62, 894.7552356719971)
                    (64, 926.4162714481354)
                    (66, 955.6197237968445)
                    (68, 1020.2898466587067)
                    (70, 1021.8034036159515)
                    (72, 1048.9826123714447)
                    (74, 1077.8257503509521)
                    (76, 1111.9297709465027)
                    (78, 1140.3197157382965)
                    (80, 1171.051645040512)
				};
				\addlegendentry{Our algorithm}
			\end{axis}
		\end{tikzpicture}
        }
		\caption{The Number of  Noises and Runtime.} 
		\label{fig:runtime}
	\end{figure}
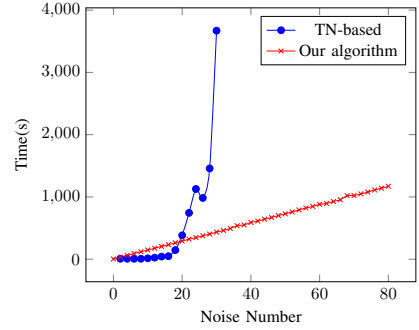

 Furthermore, to see the high efficiency of our approximation algorithm on the number of noises, we simulate the $qaoa\_100$ circuit with 0 to 80 noises as shown in Fig.~\ref{fig:runtime}. Our approximation algorithm can handle all the cases, while the TN-based method runs out of memory after the noise number reaches 30. The main reason for memory exceed is that more noise may increase the nodes' maximum rank in the tensor network contraction and consume much time and memory for contraction. As a comparison, when using level-1 approximation of Algorithm~\ref{Algorithm:Approximate}, the runtime of our approximation algorithm is almost linear with the noise number as shown in Fig.~\ref{fig:runtime}.

\begin{figure}[htbp]
    \begin{minipage}{0.24\textwidth}
        \begin{tikzpicture}[scale = 0.55]
            \begin{axis}[
                tension=0.07,
                title={Noise Rate $p=0.001$},
                xlabel={Number of Noises},
                ylabel={Sample Number},
                ymode=log,
                legend entries={{Quantum Trajectories}, {Our Algorithm}},
            ]
        \addplot[smooth,mark=*,blue] plot coordinates {
        (10, 1338.314063)
        (11, 1097.040226)
        (12, 914.294549)
        (13, 772.668911)
        (14, 660.764064)
        (15, 570.864634)
        (16, 497.599218)
        (17, 437.136068)
        (18, 386.683052)
        (19, 344.167176)
        (20, 308.023607)
        (21, 277.053469)
        (22, 250.325941)
        (23, 227.109572)
        (24, 206.823258)
        (25, 189.000717)
        (26, 173.264365)
        (27, 159.305866)
        (28, 146.871477)
        (29, 135.750875)
        (30, 125.768569)
        (31, 116.777220)
        (32, 108.652407)
        (33, 101.288497)
        (34, 94.595348)
        (35, 88.495676)
        (36, 82.922921)
        (37, 77.819516)
        (38, 73.135478)
        (39, 68.827240)
        (40, 64.856694)
        };
        \addplot[smooth,mark=x,red] plot coordinates {
        (10, 62)
        (11, 68)
        (12, 74)
        (13, 80)
        (14, 86)
        (15, 92)
        (16, 98)
        (17, 104)
        (18, 110)
        (19, 116)
        (20, 122)
        (21, 128)
        (22, 134)
        (23, 140)
        (24, 146)
        (25, 152)
        (26, 158)
        (27, 164)
        (28, 170)
        (29, 176)
        (30, 182)
        (31, 188)
        (32, 194)
        (33, 200)
        (34, 206)
        (35, 212)
        (36, 218)
        (37, 224)
        (38, 230)
        (39, 236)
        (40, 242)
        };
        \end{axis}
        \end{tikzpicture}
    \end{minipage}
    \begin{minipage}{0.24\textwidth}
        \begin{tikzpicture}[scale = 0.55]
            \begin{axis}[
                tension=0.07,
                title={Noise Rate $p=0.0001$},
                xlabel={Number of Noises},
                ylabel={Sample Number},
                ymode=log,
                legend entries={{Quantum Trajectories}, {Our Algorithm}},
            ]
        \addplot[smooth,mark=*,blue] plot coordinates {
        (10, 142878.235533)
        (11, 117986.551738)
        (12, 99062.000720)
        (13, 84340.175050)
        (14, 72663.557140)
        (15, 63247.248674)
        (16, 55543.792344)
        (17, 49161.926868)
        (18, 43816.015511)
        (19, 39293.593691)
        (20, 35433.979388)
        (21, 32113.837469)
        (22, 29237.229398)
        (23, 26728.623571)
        (24, 24527.900773)
        (25, 22586.729531)
        (26, 20865.898232)
        (27, 19333.325909)
        (28, 17962.561360)
        (29, 16731.638240)
        (30, 15622.192832)
        (31, 14618.777807)
        (32, 13708.323762)
        (33, 12879.713247)
        (34, 12123.441224)
        (35, 11431.342464)
        (36, 10796.371232)
        (37, 10212.422097)
        (38, 9674.183322)
        (39, 9177.016252)
        (40, 8716.855534)
        };
        \addplot[smooth,mark=x,red] plot coordinates {
        (10, 62)
        (11, 68)
        (12, 74)
        (13, 80)
        (14, 86)
        (15, 92)
        (16, 98)
        (17, 104)
        (18, 110)
        (19, 116)
        (20, 122)
        (21, 128)
        (22, 134)
        (23, 140)
        (24, 146)
        (25, 152)
        (26, 158)
        (27, 164)
        (28, 170)
        (29, 176)
        (30, 182)
        (31, 188)
        (32, 194)
        (33, 200)
        (34, 206)
        (35, 212)
        (36, 218)
        (37, 224)
        (38, 230)
        (39, 236)
        (40, 242)
        };
        \end{axis}
        \end{tikzpicture}
    \end{minipage}
    \caption{Comparison of Sample Number Required for the Same Error Bound}
    \label{fig:monte}
\end{figure}
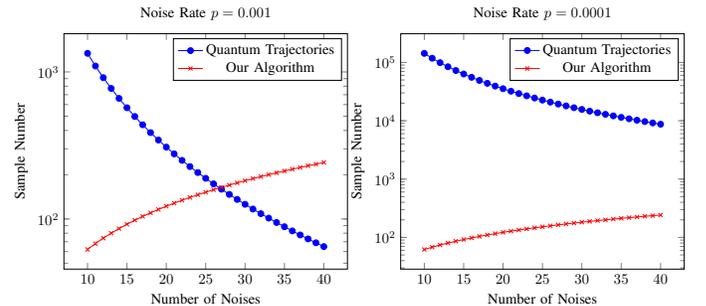

     \emph{Our Algorithm vs. Approximate Methods:} Compared to the quantum trajectories method, we fixed the success probability for it at $99\%$. We then assessed the sample numbers for our algorithm and the quantum trajectories method at different noise rates \(p\). For noise rates \(p = 0.001\) and \(p = 0.0001\), and noise number $n$ from 10 to 40, our method outperforms quantum trajectories for \(n \leq 26\) at \(p=0.001\) and consistently at \(p = 0.0001\) (See Fig.~\ref{fig:monte}).

Furthermore, we conducted numerical experiments comparing our method's efficiency and precision with various implementations of the quantum trajectories method, which is shown in Table.~\ref{table:traj_num}. The quantum trajectories method was implemented in two ways: using MM-based~\cite{isakov2021simulations} and TN-based simulator, respectively. Both tests used a depolarizing noise model with noise number 20 and rate \(p=0.001\). We adjusted the sample number for quantum trajectories to match the precision of our level-1 approximation. Seeing from Table.~\ref{table:traj_num}, our method is more efficient than the quantum trajectories method at comparable precision levels (where "Traj" abbreviates the quantum trajectories method).

\begin{table}[htbp]
\centering
\caption{\textsc{Our Algorithm v.s. Approximate Methods}}
\resizebox{\columnwidth}{!}{%
\begin{tabular}{|c|ccc|ccc|}
\hline
\multicolumn{1}{|c|}{\multirow{2}{*}{Circuit}} & \multicolumn{3}{c|}{Precision}                                                           & \multicolumn{3}{c|}{Runtime}                                                         \\ \cline{2-7} 
\multicolumn{1}{|c|}{}                         & \multicolumn{1}{c|}{Ours}      & \multicolumn{1}{c|}{Traj (MM)} & Traj (TN) & \multicolumn{1}{c|}{Ours}  & \multicolumn{1}{c|}{Traj (MM)} & {Traj (TN)} \\ \hline
QAOA\_6                                        & \multicolumn{1}{c|}{1.55E-04} & \multicolumn{1}{c|}{1.43E-04}          & 1.80E-04        & \multicolumn{1}{c|}{4.56} & \multicolumn{1}{c|}{1.49}              & 10.84           \\ \hline
QAOA\_10                                       & \multicolumn{1}{c|}{5.96E-05} & \multicolumn{1}{c|}{1.98E-04}          & 2.00E-04        & \multicolumn{1}{c|}{7.57} & \multicolumn{1}{c|}{10.94}             & 19.68           \\ \hline
QAOA\_15                                       & \multicolumn{1}{c|}{9.03E-05} & \multicolumn{1}{c|}{2.55E-04}          & 2.66E-04        & \multicolumn{1}{c|}{8.41} & \multicolumn{1}{c|}{658.38}            & 25.48           \\ \hline
QAOA\_64                                       & \multicolumn{1}{c|}{5.82E-07} & \multicolumn{1}{c|}{MO}          & 3.48E-07        & \multicolumn{1}{c|}{191.47} & \multicolumn{1}{c|}{MO}            & 598.62           \\ \hline
QAOA\_100                                       & \multicolumn{1}{c|}{1.03E-07} & \multicolumn{1}{c|}{MO}          & 1.30E-07        & \multicolumn{1}{c|}{361.79} & \multicolumn{1}{c|}{MO}            & 1264.67           \\ \hline
\end{tabular}
}
\label{table:traj_num}
\end{table}

    {\bf Analytical Experiments:} 
    We analyze our approximation algorithm in terms of noise rate and approximation levels.

\begin{figure}[tbp]
\begin{minipage}{0.24\textwidth}
    \centering
        \begin{tikzpicture}[scale = 0.55]
            \begin{axis}[
                tension=0.07,
                title={Realistic Fault Model},
                xlabel={Noise Rate},
                ylabel={Error},
            ]
        \addplot[smooth,mark=*,blue] plot coordinates {
        (0.008025188, 0.000306753)
        (0.007945336, 0.000284963)
        (0.007866279, 0.000265623)
        (0.007788008, 0.000248373)
        (0.007710516, 0.000232915)
        (0.007633795, 0.000219006)
        (0.007557837, 0.00020644)
        (0.007482636, 0.000195046)
        (0.007408182, 0.00018468)
        (0.00733447, 0.000175218)
        (0.00726149, 0.000166556)
        (0.007189237, 0.000158604)
        (0.007117703, 0.000151284)
        (0.007046881, 0.00014453)
        (0.006976763, 0.000138282)
        (0.006907343, 0.000132491)
        (0.006838614, 0.00012711)
        (0.006770569, 0.000122102)
        (0.0067032, 0.000117432)
        (0.006636503, 0.000113068)
        (0.006570468, 0.000108985)
        (0.006505091, 0.000105156)
        (0.006440364, 0.000101562)
        (0.006376282, 9.8183E-05)
        (0.006312836, 9.50009E-05)
        (0.006250023, 9.20005E-05)
        (0.006187834, 8.91678E-05)
        (0.006126264, 8.649E-05)
        (0.006065307, 8.39556E-05)
        };
        \end{axis}
        \end{tikzpicture}
\end{minipage}
\begin{minipage}{0.24\textwidth}
\centering
        \begin{tikzpicture}[scale = 0.55]
            \begin{axis}[
                tension=0.07,
                title={Depolorizing Noise Model},
                xlabel={Noise Rate},
                ylabel={Error},
            ]
        \addplot[smooth,mark=*,blue] plot coordinates {
        (0, 1.79377E-07)
        (0.0005, 2.18665E-06)
        (0.001, 8.21325E-06)
        (0.0015, 1.82591E-05)
        (0.002, 3.23243E-05)
        (0.0025, 5.04087E-05)
        (0.003, 7.25124E-05)
        (0.0035, 9.86353E-05)
        (0.004, 0.000128777)
        (0.0045, 0.000162939)
        (0.005, 0.000201119)
        (0.0055, 0.000243319)
        (0.006, 0.000289538)
        (0.0065, 0.000339776)
        (0.007, 0.000394033)
        (0.0075, 0.00045231)
        (0.008, 0.000514605)
        (0.0085, 0.00058092)
        (0.009, 0.000651254)
        (0.0095, 0.000725607)
        (0.01, 0.000803979)
        };
        \end{axis}
        \end{tikzpicture}
\end{minipage}
    \caption{Approximation error for different noise rate}
    \label{fig:decoherence_precision}
\end{figure}
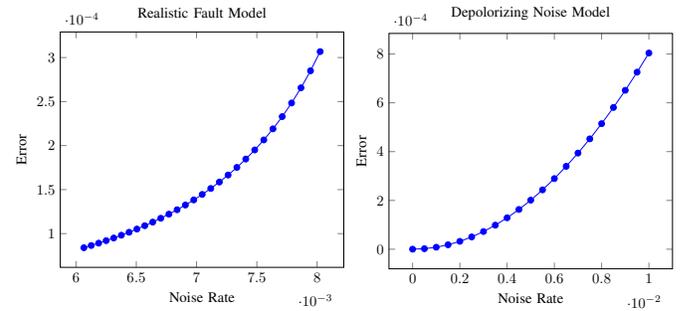

    	{\bf (1) Noise Rate:} The accuracy of our approximation algorithm depends on the noise rate of the noises, i.e., $p = \|M_{\e_D} - I\|$. When $p$ is small, the algorithm has a high accuracy. We evaluated our algorithm under both the realistic fault model and the depolarizing noise model. In both cases, the results as shown in Fig.~\ref{fig:decoherence_precision} demonstrate that as the noise rate increases, the approximation error also rises. It confirms that our approximation algorithm has better accuracy for lower noise rates in realistic noise models, which indicates a promising outlook for achieving higher precision on advanced hardware implementations in the future.

    {\bf (2) Approximation Levels:} Our algorithm offers varying levels of approximation, presenting a trade-off between computational efficiency and accuracy. As a demonstration, we simulate $qaoa\_64$ circuit with 10 noises. The input state $\ket{\psi} = \ket{0}\otimes\cdots\otimes\ket{0}$ and $\ket{v}=U\ket{0}\otimes\cdots\otimes\ket{0}$, where $U$ represents the unitary operator of the ideal circuit. Table.~\ref{table:approximate_level} shows the result for $qaoa\_64$ circuit with approximation levels from 0 to 3. From the result, we can see that the accuracy is getting higher as expected, but the cost for an extra level of approximation is also significant. For most scenarios, the level-1 approximation is recommended since it can achieve a good balance between runtime and accuracy. 

    \begin{table}[tbp]
    \centering
    \caption{\textsc{Accuracy for Different Approximation Levels}}
    \begin{tabular}{|c|c|c|c|}
    \hline
    Level    & Time (s) & Result    & Error    \\ \hline
    0        & 0.34     & 0.9539958 & 4.59E-03 \\ \hline
    1        & 11.18    & 0.9585521 & 3.02E-05 \\ \hline
    2        & 109.95   & 0.9585811 & 1.23E-06 \\ \hline
    3        & 1971.37  & 0.9585834 & 1.13E-06 \\ \hline
    \end{tabular}
    \label{table:approximate_level}
    \end{table}

 \section{Conclusion}
 
	This paper presented a novel approximation algorithm for the noisy quantum circuit simulation task. Specifically,  we introduced a new tensor network diagram for noisy quantum circuits and employed SVD to approximate the tensors representing quantum noises. In particular, our algorithm offers a speedup over the quantum trajectories method under the same approximation accuracy. Our algorithm is implemented with the Google TensorNetwork for contracting the tensor networks. The utility and effectiveness of our algorithm are demonstrated by executing noisy simulations on three types of benchmark circuits with realistic noise models. The experimental results show that our approximation algorithm can simulate the 225-qubit QAOA circuit with 20 noises. Due to the scalability of qubits,   we anticipate that our algorithm can serve as an integrated feature in the currently developed ATPG programs~(e.g.,~\cite{fang2008fault,paler2012detection,bera2017detection}) for verifying and detecting manufacturing defects, effected by quantum noises, of large-size quantum circuits.
\section*{Acknowledgments}
This work was partly supported by the Youth Innovation Promotion Association, Chinese Academy of Sciences (No. 2023116), the National Natural Science Foundation of China (Grant No. 61832015).
\printbibliography
\end{document}